\documentclass{egpubl}
\usepackage{IMET2023}
 
\WsPaper           % uncomment for final version of EG Workshop contribution
\usepackage[T1]{fontenc}
\usepackage{dfadobe}  

\usepackage{cite}
\usepackage{amsmath,amssymb,amsfonts}
\usepackage{algorithm}
\usepackage{algpseudocode}
\usepackage{graphicx}
\usepackage{textcomp}
\usepackage{xcolor}
\usepackage{tabularx}
\usepackage{multirow}
\usepackage{xurl}
\BibtexOrBiblatex
\electronicVersion
\PrintedOrElectronic
\usepackage{egweblnk} 
\title{RESenv: A Realistic Earthquake Simulation Environment
      \\based on Unreal Engine}

\author[Y. Sun, H. Wang, Z. Zhang, C. Diels, A. Asadipour]
{
\parbox{\textwidth}{
\centering 
Yitong Sun\thanks{yitong.sun@network.rca.ac.uk}\orcid{0000-0002-9469-7157}$^{1}$,
Hanchun Wang$^{2}$,
Zhejun Zhang$^{1}$,
Cyriel Diels$^{1}$,
and Ali Asadipour\thanks{ali.asadipour@rca.ac.uk}\orcid{0000-0003-0159-3090}$^{1}$
}
\\
{\parbox{\textwidth}{
\centering 
$^1$Royal College of Art\\
$^2$Imperial College London
}
}
}

\begin{document}

\maketitle

\begin{abstract}
Earthquakes have a significant impact on societies and economies, driving the need for effective search and rescue strategies. With the growing role of AI and robotics in these operations, high-quality synthetic visual data becomes crucial. Current simulation methods, mostly focusing on single building damages, often fail to provide realistic visuals for complex urban settings. To bridge this gap, we introduce an innovative earthquake simulation system using the Chaos Physics System in Unreal Engine. Our approach aims to offer detailed and realistic visual simulations essential for AI and robotic training in rescue missions. By integrating real seismic waveform data, we enhance the authenticity and relevance of our simulations, ensuring they closely mirror real-world earthquake scenarios. Leveraging the advanced capabilities of Unreal Engine, our system delivers not only high-quality visualisations but also real-time dynamic interactions, making the simulated environments more immersive and responsive. By providing advanced renderings, accurate physical interactions, and comprehensive geological movements, our solution outperforms traditional methods in efficiency and user experience. Our simulation environment stands out in its detail and realism, making it a valuable tool for AI tasks such as path planning and image recognition related to earthquake responses. We validate our approach through three AI-based tasks: similarity detection, path planning, and image segmentation.

\begin{CCSXML}
<ccs2012>
<concept>
<concept_id>10010147.10010341.10010366.10010367</concept_id>
<concept_desc>Computing methodologies~Simulation environments</concept_desc>
<concept_significance>500</concept_significance>
</concept>
<concept>
<concept_id>10010147.10010178.10010224</concept_id>
<concept_desc>Computing methodologies~Computer vision</concept_desc>
<concept_significance>500</concept_significance>
</concept>
<concept>
<concept_id>10003120.10003145.10003151.10011771</concept_id>
<concept_desc>Human-centered computing~Visualization toolkits</concept_desc>
<concept_significance>500</concept_significance>
</concept>
</ccs2012>
\end{CCSXML}

\ccsdesc[500]{Computing methodologies~Simulation environments}
\ccsdesc[500]{Computing methodologies~Computer vision}
\ccsdesc[500]{Human-centered computing~Visualization toolkits}

\printccsdesc   
\end{abstract}  
%-------------------------------------------------------------------------
\section{Introduction}
Earthquakes, as recurrent natural disasters, profoundly impact human life and economic activities \cite{joseph2022effect}. To mitigate the adverse effects and enhance the efficiency of post-earthquake rescue operations, a new paradigm shift towards the integration of artificial intelligence (AI) and robotics has been observed \cite{nazarova2020application}. These technologies, encompassing tasks such as path planning, automatic obstacle avoidance, and image recognition, bring about significant improvements in earthquake response activities \cite{magid2019artificial}. Yet, given the unpredictable and dynamic nature of earthquakes, creating a platform to iteratively train these technologies across a myriad of realistic earthquake scenarios becomes a pressing need.

While virtual game engines have emerged as a potent tool for simulating various disaster situations, like firestorms or flash floods \cite{bourhim2020selection}, their application for earthquake scenarios, despite their potential, remains largely unexplored. These engines not only offer reduced operational complexity but also ensure shorter and more efficient training cycles than their traditional counterparts. Furthermore, the integration of cutting-edge realistic rendering techniques, especially ray tracing, is paving the way for substituting real-world data, thereby revolutionising visual recognition research \cite{greff2022kubric}. However, harnessing these advanced techniques and tools specifically for earthquake simulations within a virtual environment has been a gap in the current research landscape.

To address this gap, we present RESenv—an environment tailored for earthquake simulation, leveraging the capabilities of the Chaos physics engine within Unreal Engine 5 (UE5). By using actual seismic waveform data sourced from online repositories and integrating it into UE5, RESenv aims to offer a nuanced simulation of building destruction during earthquakes. The overarching goal of this novel environment is to furnish high-resolution, in-depth visual, and scenario simulations. This not only aids in search and rescue mission planning but also emerges as an invaluable synthetic data reservoir for AI training in diverse applications such as path planning and visual recognition.

Our key contributions in this study include:

\begin{enumerate}
    \item We introduce a method based on UE5, aiming to provide a realistic earthquake simulation. This approach simulates multi-building scenarios by capitalising on actual seismic data sourced online.
    
    \item The endorsement of the applicability and efficiency of RESenv via three AI-centric tasks, namely similarity detection, path planning, and image segmentation. This solidifies RESenv's potential as an exemplary environment that equips AI models with a rich dataset.
    
    \item A paradigmatic shift from conventional methods to a more streamlined, intuitive, and accessible approach, courtesy of the game engine's versatility, effectively dismantling domain-specific barriers and democratising access.
\end{enumerate}

\section{\textbf{Related Work}}
This section reviews research related to earthquake simulation and AI training for rescue missions, which form the basis for our proposed UE-based earthquake simulation approach.

\subsection{\textbf{Earthquake Simulation}}
Earthquake simulation, a longstanding research focus in geophysics, geology, and engineering \cite{matin2022challenges}, has seen significant advancements due to recent progress in computer hardware. This has enabled more sophisticated modelling of earthquakes and consequent building damage using numerical simulation techniques \cite{xu2020photo, shaw2022earthquake}. Stress simulations of individual buildings, initially aimed at analysing seismic stress-induced deformation and structural optimisation, have matured, and some researchers have employed finite element analysis (FEA) for assessing earthquake-induced building damage and exploring risk mitigation strategies \cite{xu2018post, mckenna2011opensees}. However, due to real-world buildings' structural complexity, material diversity, and computational constraints, most simulations only model primary load-bearing structures and facades, resulting in discrepancies between simulated and actual outcomes. Current earthquake platforms, constrained by the complexity of the physics engine limits and simulating only single or two degree-of-freedom (DOF) vibrations, fail to mimic the three DOF motions of actual earthquakes \cite{Simpson, oleson}.

Urban multi-building simulations, compared to individual building simulations, emerged much later. One of the most widely used frameworks is HAZUS, developed by the United States Federal Emergency Management Agency (FEMA) \cite{schneider2006hazus}. Based on standardised Geographic Information System (GIS) methodologies, HAZUS is employed for estimating the impact of earthquakes, post-earthquake fires, floods, and hurricanes, among other disasters. To overcome the limitations of HAZUS's single DOF model, Japanese researchers introduced the Integrated Earthquake Simulation (IES) framework, utilising multi-dimensional data fusion calculation methods \cite{hori2011introduction}. Subsequently, Turkish researchers developed a regional building simulation method for Istanbul using MATLAB, based on the IES framework \cite{sahin2016development}. Similar to individual building simulations, multi-building simulations are also constrained by software limitations in terms of physical collisions and building structural complexity. Although a study by David et al. employed large-scale computing to simulate the motion of geological faults and measure building responses \cite{mccallen2021eqsim}, the focus of these research predominantly lies in calculating the complexity of geological structures, with scant attention paid to the fidelity of building structures.

In our approach, we utilise the Chaos destruction system (\webLink{https://docs.unrealengine.com/5.1/en-US/destruction-overview}) and Nanite visualisation system (\webLink{https://docs.unrealengine.com/5.1/en-US/nanite-virtualized-geometry-in-unreal-engine}) within the UE 5 game engine, achieving previously challenging fracture and fragmentation simulations for different materials and complex hybrid structures, accurate physical collisions, and three DOF geological motion. Our method demonstrates a significant improvement in computational efficiency and cost compared to conventional techniques, enabling real-time and accelerated calculations on consumer-grade computers. Benefiting from the user-friendliness of the game engine, we have established a pre-fabricated library of materials and structures readily available to users by creating pre-set programs, greatly simplifying operational difficulties compared to traditional simulation frameworks, and thus enabling researchers from various fields to use our approach with ease.

\subsection{\textbf{AI Training for Rescue Missions}}
AI applications in search and rescue operations, as well as the robotics domain, are becoming increasingly widespread. Numerous researchers are dedicated to employing deep learning and reinforcement learning techniques for complex terrain path planning, image recognition, and other related tasks \cite{costa2019survey, queralta2020collaborative}. For instance, the study by LinLin et al. utilised the SBMPC algorithm to investigate path planning problems for search and rescue robots \cite{wang2020research}. Xuexi et al. explored indoor search and rescue using Simultaneous Localisation and Mapping (SLAM) and Light Detection and Ranging (LiDAR) methods \cite{zhang20202d}. Farzad's research introduced the application of deep reinforcement learning (DRL) methods in search and rescue robot tasks \cite{niroui2019deep}. These studies underscore the potential of artificial intelligence in enhancing the effectiveness and efficiency of search and rescue missions.

The success of AI approaches largely depends on the availability of ample and high-quality data as training inputs, which can accurately represent the complexities and dynamics of environments affected by earthquakes \cite{bischke2019multimedia}. However, in earthquake rescue scenarios, collecting and obtaining real-world data poses significant challenges. To address data limitations, researchers have developed various virtual environments, such as the RoboCup Rescue Simulation Environment \cite{skinner2010robocup}, USARSim \cite{polverari2007development}, and the BCB environment developed by Laurea University of Applied Sciences \cite{grunwald2018reliability}, for creating training data for deep learning and reinforcement learning algorithms in search and rescue operations. Nonetheless, these frameworks currently lack the level of texture rendering realism and detail richness required for training AI models that rely on image recognition and depth data inputs. This also results in substantial discrepancies between the volume and complexity of simulated scenarios and actual search and rescue missions.

\begin{figure*}[t!]
 \centering
 \includegraphics[width=\textwidth]{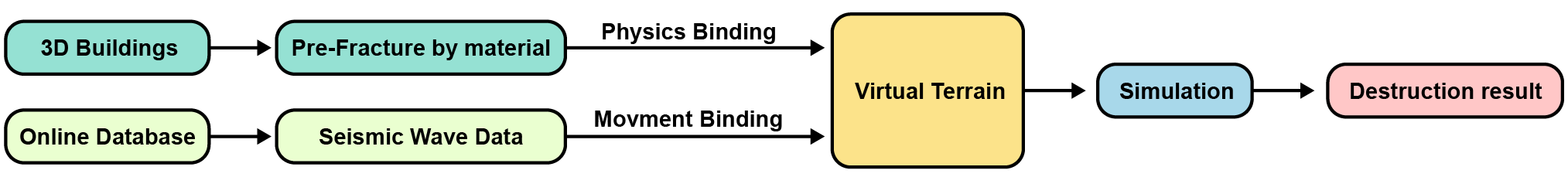}
 \caption{Flowchart of RESenv for earthquake simulation. There are three steps: scenario preparation, data binding, and simulation. During the data preparation phase, 3D building models are imported into UE, which are then pre-fractured by material groups. Actual seismic wave data was acquired from the IRIS online database and imported via a graphical UI system. In the data binding phase, the 3D buildings are bound to the virtual terrain by PCA through an automated analysis program. The seismic wave data is converted into terrain displacements for binding. During the simulation phase, RESenv is run at a specified frame rate, 40 FPS for desktop computer simulation, 90 FPS for VR training, and 240 FPS for high frame rate sensor training. As the simulation begins, the seismic wave data displaces the terrain, which in turn causes the pre-fractured 3D buildings to be destroyed. RESenv is interactable throughout the simulation.}
 \label{fig:process}
\end{figure*}

Our proposed simulation environment fills this void by aiming to provide a highly realistic and detailed virtual earthquake damage environment using a ray-tracing system and authentic scanned textures. The environment allows for generating high-quality training data that can be directly used for AI algorithm visual recognition and depth data scanning. Weather phenomena, lighting conditions, and post-earthquake dust will be effectively simulated.

\section{\textbf{Method}}
RESenv executes the earthquake simulation in a three-stage process: \textbf{data preparation}, \textbf{data binding}, and \textbf{simulation}. Figure \ref{fig:process} describes the flowchart of the method.

\subsection{\textbf{Data Preparing}}
\textbf{Virtual Building Processing} \hspace{0.2cm} Due to the flexibility in model importation within UE, virtual building models based on Polygon Mesh can be acquired from various sources. For instance, manual creation using modelling software like Blender, computation from GIS data in CityEngine software \cite{badwi20223d}, or generation via AI methods \cite{chaillou2019ai}. However, to ensure that the building models can be effectively simulated, the models first need to be pre-processed before importing them into our method through UE. Initially, the size units of the models need to be standardised. Typically, Polygon Mesh-based models do not have a unified scale unit like NURBS Surface models; the models need to be pre-scale to align to the cross-platform unified units. In RESenv, the default length unit in UE5 is adopted (centimetre) . Secondly, building models need to be segmented according to distinct materials, such as concrete, bricks, and wooden structures, to set up pre-fracture settings (\webLink{https://docs.unrealengine.com/5.1/en-US/destruction-overview/}) separately after importing into UE.

\begin{figure}[!b]
 \centering
 \includegraphics[width=\columnwidth]{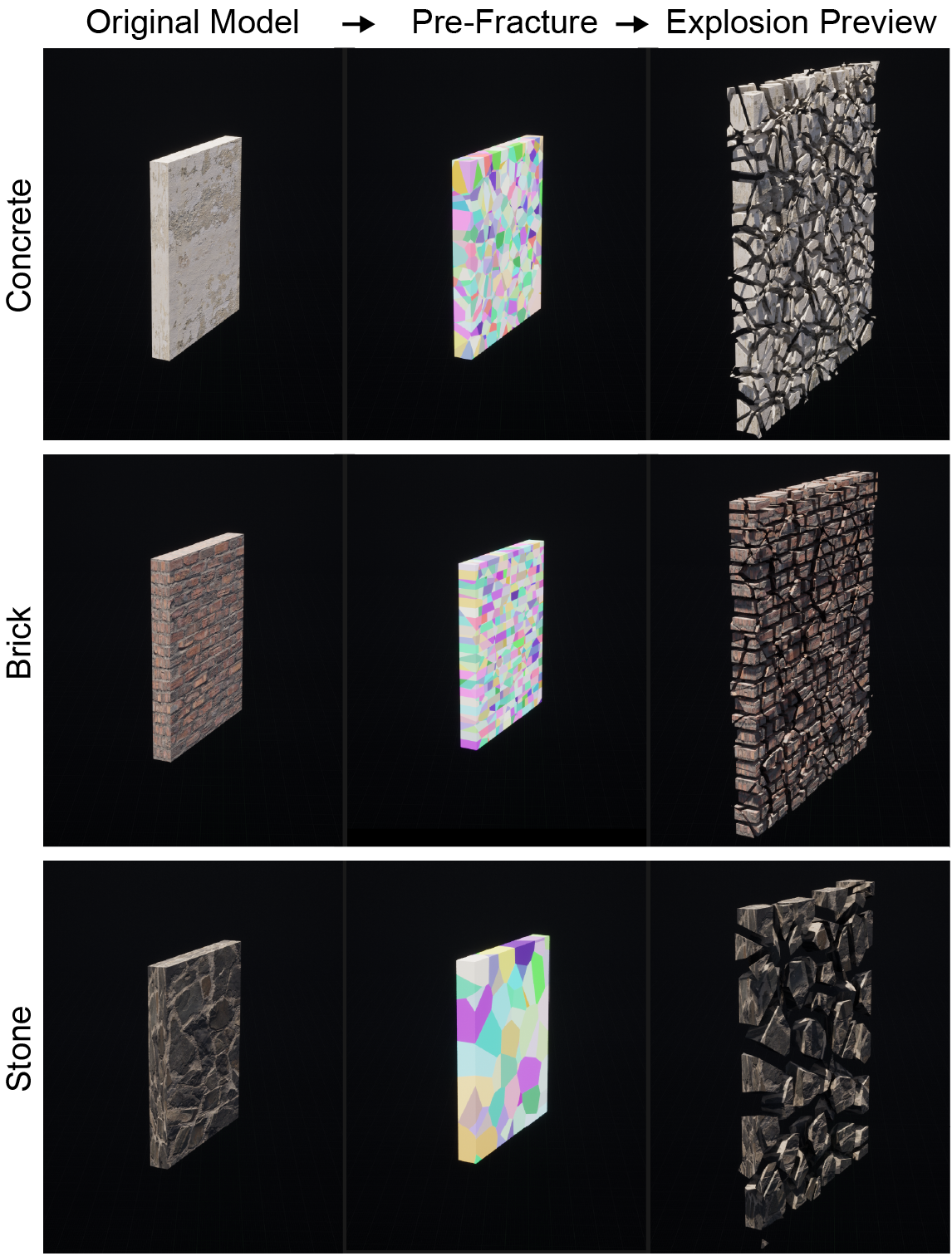}
 \caption{Three examples of pre-fracturing of walls in different materials. Once the building has been imported into UE, the 3D model needs to be pre-fractured in order to be damaged accordingly in the earthquake simulation. The pre-fracture is set according to the material properties of the building. The three instances show the set-up process, segmentation methods, and explosion view for concrete, brick wall, and stone structures, respectively.}
 \label{fig:threeMaterial}
\end{figure}
 
In the process of calibrating our pre-fracture settings, the material mechanics data published by MatWeb (\webLink{https://www.matweb.com}) and Material Data Repository (\webLink{https://materialsdata.nist.gov}) served as crucial references. Specifically, we employed Ansys to incorporate this data, setting up analogous conditions within UE to discern the optimal pre-fracture parameters. By aligning the mechanical properties derived from these databases with UE's pre-fracturing capabilities, we ensured that our simulations were as realistic and accurate as possible. Additionally, to facilitate usability, we have meticulously curated a comprehensive material library, with plans for its continuous expansion. For users inclined towards customising the pre-fracture settings further, we recommend initially consulting our material library settings as a baseline, and subsequently referring to the official Unreal Engine documentation for tailored adjustments and configurations. This strategy not only ensures that our method remains grounded in validated material science but also offers flexibility for advanced users to fine-tune their simulations.

\begin{figure}[!t]
 \centering
 \includegraphics[width=\columnwidth]{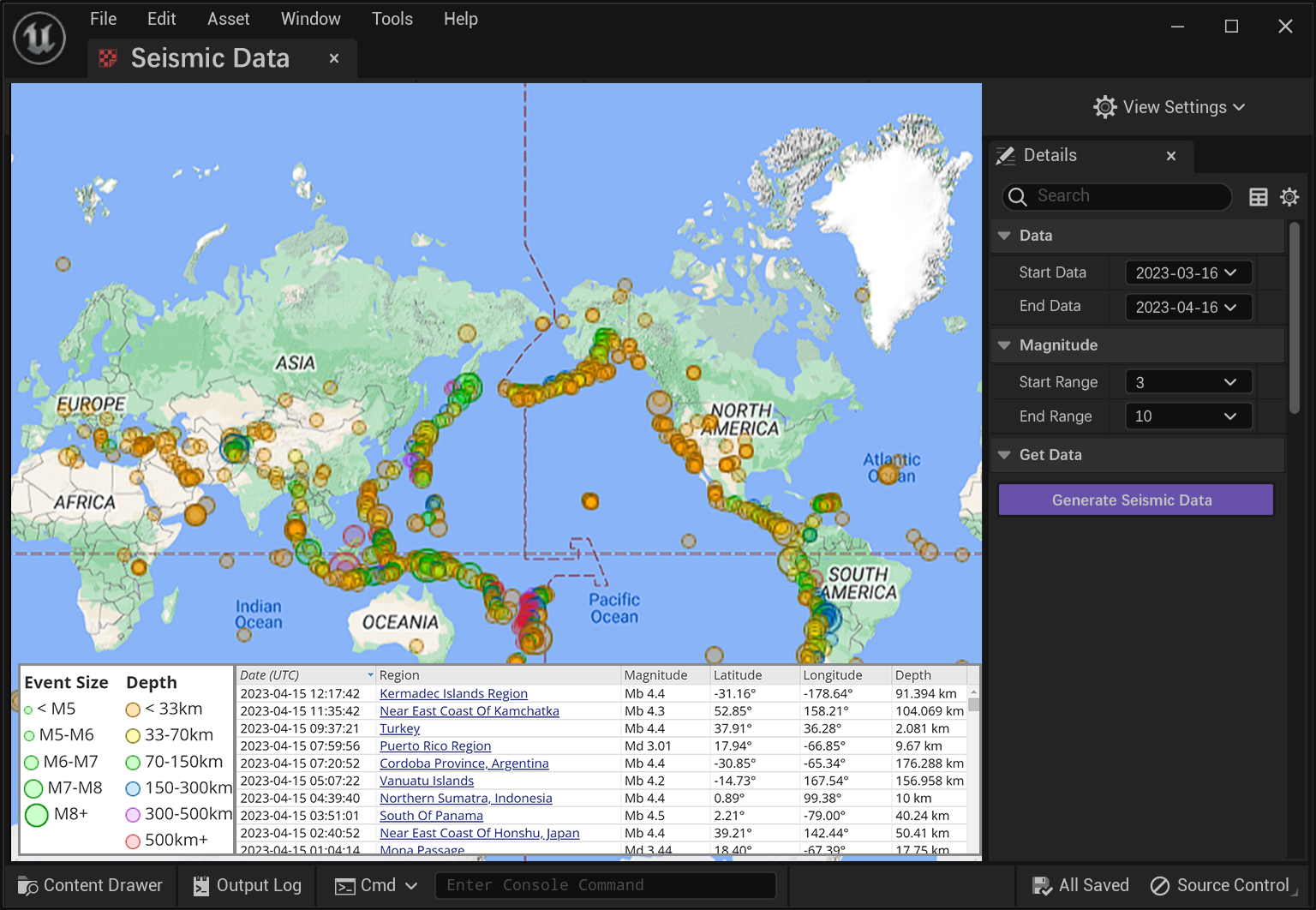}
 \caption{The RESenv user interface for acquiring IRIS online seismic data. The user interface contains an interactive world map that can be clicked on to select the seismic data to be acquired. The column on the right side allows to define time ranges and earthquake levels as a filter.}
 \label{fig:UI}
\end{figure}

In contrast to traditional earthquake simulation platforms that only simulate the physical collision between blocks and joints, our approach based on the Chaos Physics System in UE5 can achieve effects similar to real building destruction. Figure \ref{fig:threeMaterial} presents examples of pre-fracture settings for three different wall materials. As seen in the figure, different segmentation strategies and levels can be applied according to distinct materials. The fractured geometry collection can be set with different break-force thresholds to represent the strength of the materials.

\textbf{Seismic Wave Data Acquisition} \hspace{0.2cm} The seismic waveforms used by earthquake simulation can be divided into two categories. 1) Waveforms recorded from actual earthquakes that have already occurred. These waveforms can be obtained from publicly available datasets online. An earthquake event is often recorded by multiple seismometers located at different geographical locations; by cross-comparing and applying algorithms for noise reduction, the absolute motion of the Earth's surface can be authentically reproduced in simulation platforms. 2) Waveforms synthesised through algorithms \cite{moseley2018fast, moseley2020deep}. In seismic resistance testing of buildings, researchers have developed various methods to synthesise earthquake waveforms in order to assess the impact of different levels and types of earthquakes on building structures. This enables the simulation platform to carry out unlimited iterations of earthquake tests in any conditions. Our method primarily aims to simulate the damage sustained by urban buildings in actual earthquakes to provide realistic datasets for AI visual-based training; therefore, the method initially implements the simulation of global earthquake waveforms obtained from the IRIS online database (\webLink{https://ds.iris.edu/ds/nodes/dmc/data/}). The acquired seismic waveform data records three DOF of geological movement, named "BH1" (east-west direction), "BH2" (north-south direction), and "BHZ" (vertical direction). We have implemented a user-friendly user interface (UI) and Python-based automatic format conversion program in UE (Figure \ref{fig:UI}), enabling users to directly obtain seismic waveforms by clicking on the geographical location and event occurrence time on the global map without the need for complex data searches and imports. Once the user selects the required data, the waveforms are automatically converted into a "DataTable" file supported by UE.

\begin{figure}[t]
\centering
\includegraphics[width=\columnwidth]{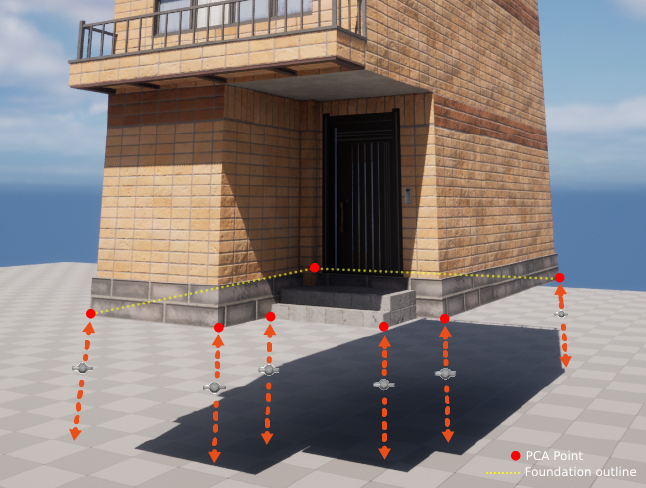}
\caption{An exemplification of a 3D building, autonomously anchored to a virtual terrain through the PCA placement program.}
\label{fig:binding}
\end{figure}

\subsection{\textbf{Data Binding with Virtual Terrain}}

In the real world, structures are anchored to the ground, responding dynamically to forces from seismic activities. Mimicking this, our method endeavours to realistically simulate the anchor forces exerted on buildings during ground movements. Consequently, the RESenv necessitates the presence of a virtual terrain within UE to serve as the anchoring ground, binding the movements generated by the earthquake data.

The choice of using the "Physics Constraint Actor" (PCA) (\webLink{https://docs.unrealengine.com/5.1/en-US/physics-constraints-in-unreal-engine/}) in the UE physics system for binding buildings to the virtual terrain is motivated by its robust capability to simulate real-world constraint forces. It accurately mirrors the building-ground interaction during earthquakes based on their physical attributes. This binding mechanism establishes a series of physical anchor force constraints between the structure and the terrain, wherein the building's linear motion threshold and rotational thresholds within the Cartesian coordinate system are determined by the building's material properties and volume.

\begin{figure*}[t]
 \centering
 \includegraphics[width=\textwidth]{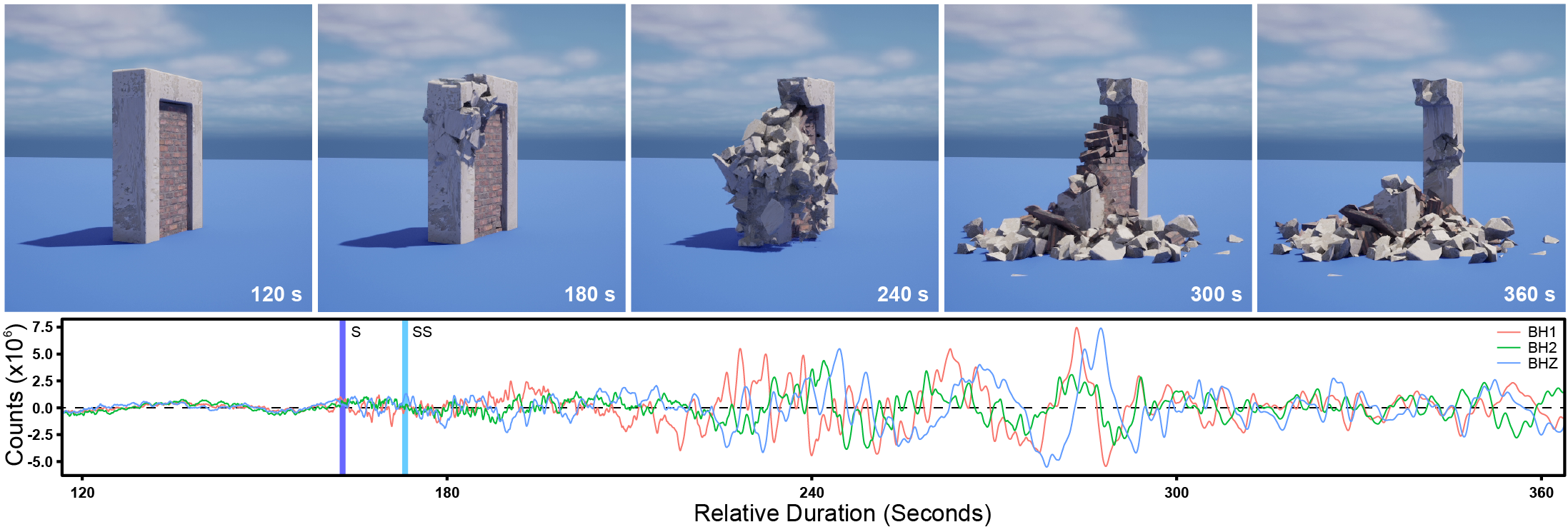}
 \caption{An example showing the process of earthquake simulation by RESenv. In this example, a wall with a concrete and brick structure is pre-fractured and bound to a virtual terrain. The data taken from the 7.4 magnitude earthquake in Oaxaca, Mexico, on 23 June 2020, recorded by the seismic station code-named TEIG. The entire simulation lasted 360 seconds. Five frames were extracted to demonstrate wall damage changes during the simulation.}
 \label{fig:duration}
\end{figure*}

A cornerstone of our approach is the foundation weight map, which is derived from the building's three-dimensional volume. Larger volume regions bear higher weight, ensuring that the PCA's anchoring force corresponds authentically to the actual size and shape of the building. For instance, during minor seismic activities, the building's foundation and walls will undergo nominal displacements without detaching from the terrain. However, during more intense quakes, the building components will fracture and potentially collapse, aligning with the intrinsic architectural design.

One salient challenge in placing the PCAs is preventing an overwhelming concentration of constraint forces within a singular area, which might result in unrealistic building movements. Meanwhile, too dense constraint placement will cause serious waste of computing resources. To surmount this, we've devised an algorithm that dynamically identifies optimal PCA positions based on the foundation's contours and inflection points. Furthermore, the RESenv deploys a C++ program to automate the analysis of the building foundation's shape, primarily its planar geometry. By identifying edge transitions, we systematically place the PCA while ensuring a minimum distance between them. Appendix B lists pseudocode for performing PCA binding. Figure \ref{fig:binding} demonstrates the distribution of binding points when a building is automatically bound to the ground in our approach.

To facilitate the simulation of terrains moving akin to the Earth's crust during seismic events, we've harnessed a C++ program. This program associates the previously extracted earthquake waveform data - "BH1", "BH2", and "BHZ" - with the "X", "Y", and "Z" axes, governing the terrain's motion. Earthquake waveform data from the IRIS database exhibits a frequency of 40 Hz, implying 40 recorded samples every second. In our simulation, RESenv offers three distinct frame rates in UE: 40 FPS, 90 FPS, and 240 FPS. These rates cater to different applications, including desktop simulations, VR training, and high frame rate sensor data synthesis. For handling the data at 90 and 240 FPS rates, we employed the wavelet interpolation algorithm, a method renowned for its ability to preserve signal details while increasing the sample rate \cite{yu2007wavelet}.

\subsection{\textbf{Simulation}}
Upon completion of the data binding, our method can be executed in UE in Simulate In Editor (SIE) mode (\webLink{https://docs.unrealengine.com/5.1/en-US/in-editor-testing-play-and-simulate-in-unreal-engine}). It is worth noting that, unlike traditional simulation approaches, our method inherits features from Unreal Engine, allowing all virtual assets and fractured models to be interactive during run-time. Applications such as VR search and rescue training and robotic dynamic obstacle avoidance will transition from static scene training to dynamic training with time-varying properties. Figure \ref{fig:duration} displays an example of a complete simulation process. In this instance, a concrete frame and a brick wall are bound to a flat virtual ground. The seismic data is sourced from the magnitude 7.4 earthquake in Oaxaca, Mexico, on June 23, 2020, recorded by the seismograph station coded as TEIG. The simulation lasts 360 seconds and runs at a frame rate of 40 FPS.

\section{\textbf{Experiment}}
In order to verify the efficacy of our approach, two earthquake simulation experiments with three tasks were designed with the aim of providing synthetic visual data for AI training. 1) Two historical earthquake events and two laboratory experiments were reproduced and simulated with damage to buildings with four different materials. The similarity between real and simulated damaged buildings was assessed using a pre-trained Vision Transformer (ViT) model.  2) Using GIS data, we recreated a Japanese neighbourhood and then conducted a 20 random endpoints robot path planning test in the simulated post-earthquake area based on synthetic visual data. The completion rates of the robot's path and the success rates of visual recognition en route are counted.

\begin{figure*}[t!]
 \centering
 \includegraphics[width=0.8\textwidth]{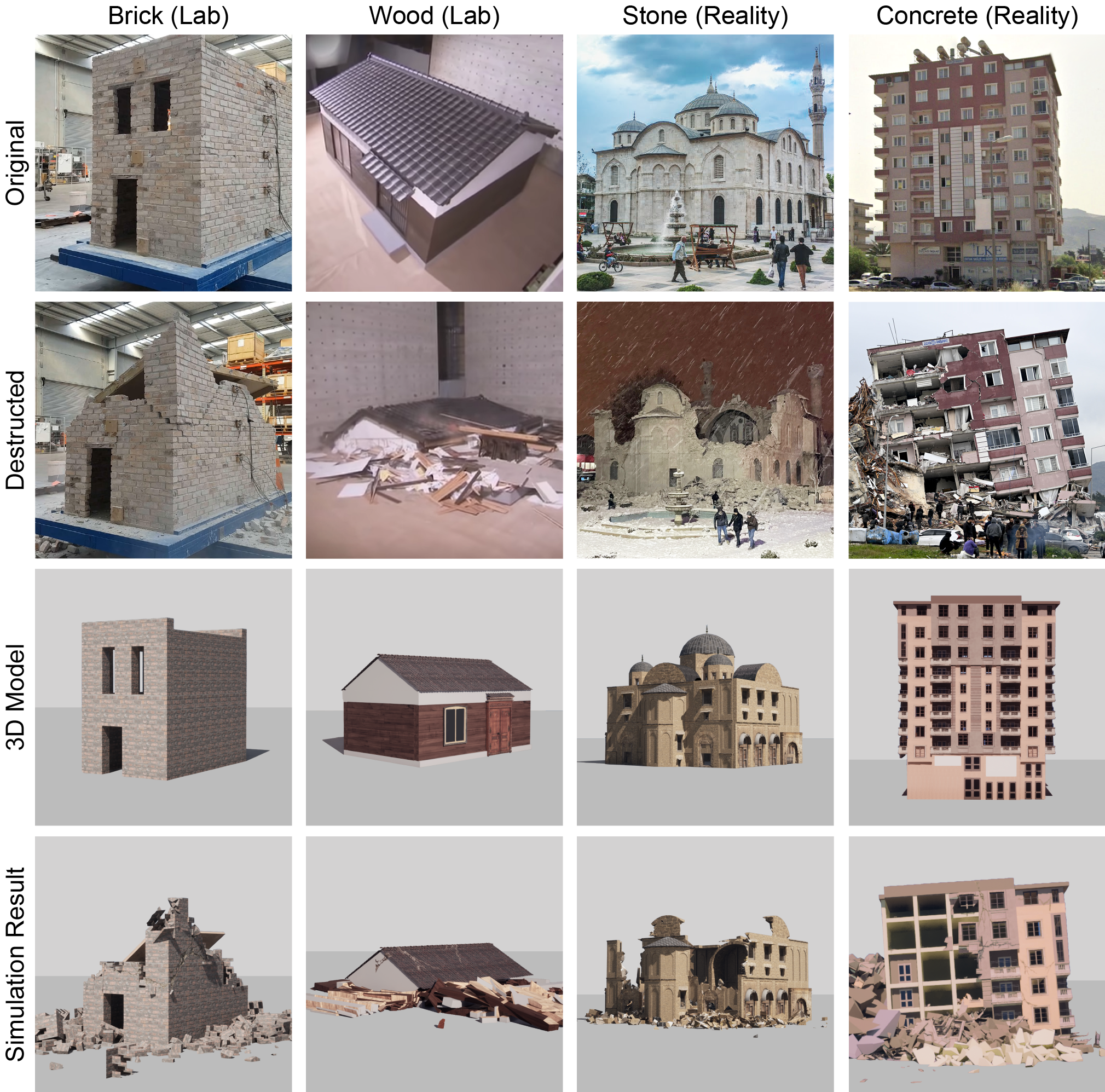}
 \caption{Earthquake simulation using RESenv for four actual scenarios. Two laboratory experiments (columns 1-2) and two actual buildings (columns 3-4) were chosen. Row 1: The original forms of four buildings before the earthquake events. Row 2: Destruction of buildings following the earthquake events. Row 3: The 3D models are recreated based on the actual buildings and are given textures and pre-fracture settings in RESenv. Row 4: Destruction results of four buildings after RESenv earthquake simulation.}
 \label{fig:fourScene}
\end{figure*}

\subsection{\textbf{Realistic Simulation of Buildings}}
Four real-world earthquake-induced building damage scenarios are selected and re-created the destruction in a simulated environment using RESenv to compare the accuracy of our approach. Figure \ref{fig:fourScene} presents the original references and simulation results of the four scenarios. Each scenario simulates a building structure based on different materials. Two were simulated on a laboratory shake table \cite{giaretton2022shaking, noauthor_6_nodate}, and two were natural earthquakes \cite{michaelson_earthquake_2023, noauthor_turkey_2023}.

We obtained the building data and seismic wave data of the two laboratory-simulated scenarios via email correspondence. The seismic wave data of the two real-world scenarios were directly obtained from the IRIS database through the RESenv UI system, while the two buildings were reconstructed from multiple viewpoints using multiple online references. All four buildings were constructed in Blender, and surface textures were obtained from the Quixel Megascans material library (\webLink{https://quixel.com/megascans}). The simulation was carried out in UE 5.1.1 on a Razor laptop with an RTX-3070 GPU, an AMD Ryzen 6900HX CPU, and 16GB of RAM as the minimal requirement. The initial UE scenario was set to the default configuration.

To address the challenges arising from the inherent complexities of real-world building materials and structures, which are difficult to replicate perfectly in simulations, we adopted a more exhaustive approach for similarity detection. Recognising that a single simulation might not capture the full extent of potential damages, we conducted 100 random pre-fracture simulations for each building scenario, ensuring consistent material properties across iterations. Subsequent to this, iterative similarity checks were performed using our algorithm. The results presented are based on the highest similarity scores obtained, which, we believe, best represent the capabilities of our proposed method. This strategy not only helps in capturing the variability in damage patterns but also underscores the robustness of our approach in achieving close resemblance despite the inherent unpredictability and complexities of real-world scenarios.

To address the challenges arising from the inherent complexities of real-world building materials and structures, such as material ageing, non-uniformity, and construction errors, we adopted a more exhaustive approach for similarity detection. Recognising that a single simulation might not capture the full extent of potential damages, we conducted 100 random pre-fracture simulations for each building scenario to approximate the properties of real building materials iteratively. Following this exhaustive pre-fracture simulation, to ascertain the similarity between the simulated results and actual structural damage — especially in the context of robotic visual recognition tasks — we employed a verified ViT deep learning model designed for feature similarity assessment. This model was pre-trained on the widely-used ImageNet-21K dataset, and its efficacy has been validated in research by Omini et al. \cite{omori2022predict}. Images of real-world damaged buildings and those from our simulations were independently input into the ViT model to compute their similarity. The results, based on the highest similarity scores achieved, are believed to best represent the capabilities of our proposed method. This strategy not only captures the variability in damage patterns but also underscores our approach's robustness in mimicking real-world scenarios despite their inherent unpredictability and complexities. The final computational outcomes are presented in Table \ref{tab1}.

\begin{table}[htbp]
\caption{Similarity of the four scenarios simulated with RESenv over 100 attempts, with the best simulation run indicated.}
\begin{center}
\renewcommand{\arraystretch}{1}
\begin{tabularx}{\linewidth}{c X X X X }
\hline 
\textbf{Scenario} & \textbf{Brick} & \textbf{Wood} & \textbf{Stone} & \textbf{Concrete}\\
\hline
\textbf{Similarity} & 94.80\% & 90.07\% & 92.25\% & 89.34\%\\

\textbf{Best Attempt} & 57 /100 & 83 /100 & 22 /100 & 48 /100\\
\hline
%\multicolumn{5}{l}{$^{\mathrm{a}}$Sample of a Table footnote.}
\end{tabularx}
\label{tab1}
\end{center}
\end{table}

% Similarity detection between real building damage and RESenv simulation results was achieved by using the pre-trained ResNet-50 model as a vector feature extractor and the Siamese network as a feature comparator.

The results demonstrate that the simulations for all four structures attained a considerable degree of resemblance to real-world scenarios. Our proposed earthquake simulation technique exhibits a robust capability to accurately replicate the damage patterns induced by actual seismic events in buildings. The observed discrepancies in the outcomes could be attributed to variations in the pre-fracture parameters of the 3D building materials, as compared to those of the reference structures. Consequently, these disparities give rise to differences in the morphology and movement trajectories of the fragmented masses within the simulation.

\begin{figure*}[t!]
 \centering
 \includegraphics[width=\textwidth]{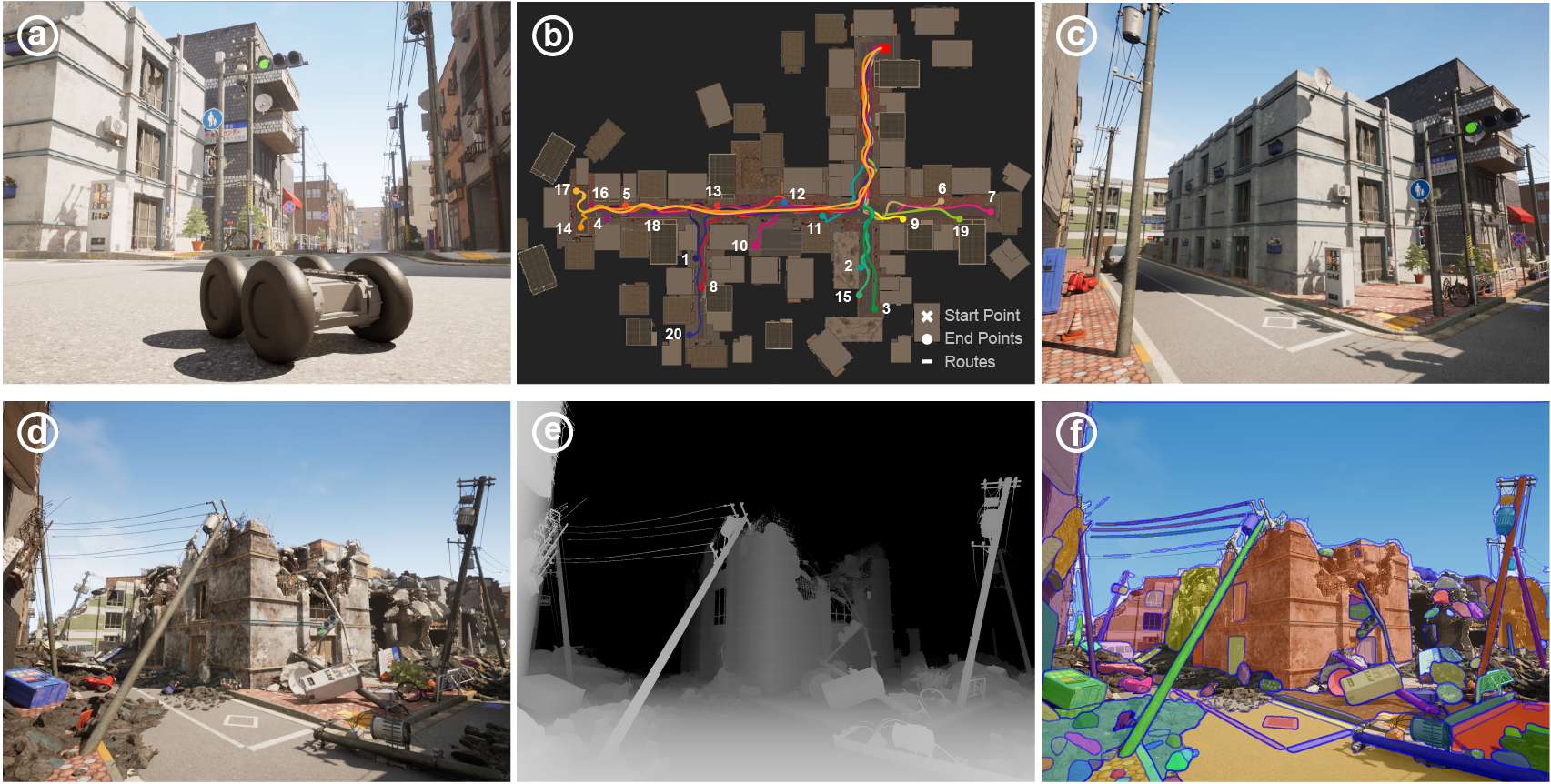}
 \caption{Multi-building scenario earthquake simulation experiments in RESenv. Task 1 is to perform a robot path planning and obstacle avoidance test using the pre-trained DRL SLAM to verify the effectiveness of the earthquake simulation scenario for robot training. Task 2 is to use SAM to perform image segmentation detection on synthetic data from a RGB camera while the robot is travelling. Ultimately, the segmentation success rate will be counted. a: buildings generated in CityEngine using GIS data. A rover robot is placed in the scenario to perform path-planning tasks based on SLAM DRL. b: A record of 20 randomly selected endpoints for the path planning task. c: simulated RGB camera view of the original scenario in UE. d: RGB camera view of the scenario after an earthquake simulation. e: simulated depth camera view for SLAM DRL algorithm data input. f: RGB camera view with SAM image segmentation.}
 \label{fig:experiment}
\end{figure*}

\subsection{\textbf{Scenario Simulation and Robotic Training}}
\begin{figure*}[htbp]
 \centering
 \includegraphics[width=0.9\textwidth]{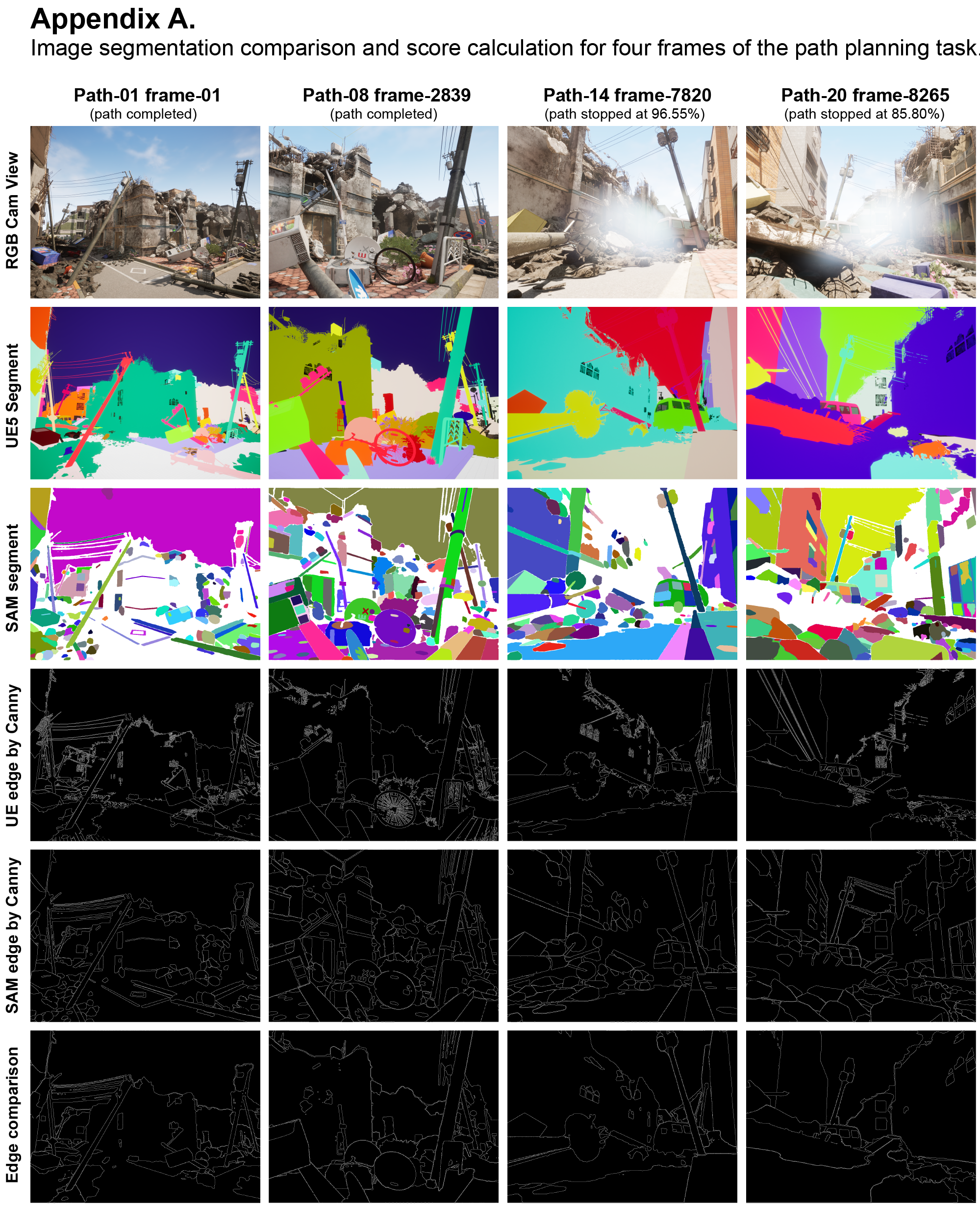}
 \caption{Comparison of image segmentation and score calculation for 4 typical paths. \textbf{Row 1:} Synthetic RGB camera shot. \textbf{Row 2:} Object image segmentation using UE5 built-in camera post-processing material (ground truth). \textbf{Row 3:} Image segmentation of the RGB shot image using the SAM model. \textbf{Row 4:} UE segmentation edge map calculated using the Canny algorithm with edge dilation. \textbf{Row 5:} SAM segmentation edge map calculated using the Canny algorithm with edge dilation. \textbf{Row 6:} The segmentation score was calculated as the proportion of overlapping edge pixels to the total edge pixels in the UE segmentation.}
 \label{fig:appendix_01}
\end{figure*}

To evaluate the effectiveness of RESenv in conducting earthquake simulations within urban settings featuring clusters of buildings for robotic training, two distinct tasks are designed. Initially, the GIS data from a Japanese neighbourhood were obtained via OpenStreetMap and subsequently converted into a 3D scene utilising CityEngine. The scene was then imported into UE. All buildings within the scene were automatically anchored to the terrain using the program in RESenv, while the terrain was linked to earthquake data. A robot model, sourced from RoverRobotics\textsuperscript{\textcircled{R}}, was positioned in the scene and equipped with simulated RGB and depth camera sensors (Figure \ref{fig:experiment}.a, d, e). The robot was assigned two tasks: 1) Utilising a DRL model based on SLAM, as proposed by Shuhuan et al. \cite{wen2020path}, the robot was instructed to randomly select 20 coordinates as endpoints for path planning and obstacle avoidance testing within the simulated environment (Figure \ref{fig:experiment}.b). The ratio of the completed length of each path to its total length is recorded. This test aimed to verify whether our simulation framework could provide effective earthquake scenarios for AI path-planning methods with demonstrated efficacy. 2) Concurrently with Task 1, the Segment-Anything Model (SAM)(model: ViT-H)\cite{kirillov2023segment} was selected as the state-of-the-art for generalised image segmentation model to collect data from RGB sensors for object segmentation (Figure \ref{fig:experiment}.f). Image segmentation and its edge detection serve as the foundation for training AI models and path-planning tasks. For each frame from the camera, we compared the alignment of edges between the built-in UE segmentation (ground truth), SAM-processed ground truth, and SAM segmentation (Appendix A fig.\ref{fig:appendix_01}. Row 2-3). By using the Canny algorithm with edge dilation for pixel-level alignment deviation, the segmentation score was calculated as the proportion of overlapping edge pixels to the total edge pixels in the UE segmentation (Appendix A fig.\ref{fig:appendix_01}. Row 4-6). The final success rate for each path is derived from the average of all frames.  

Upon completing the aforementioned tasks, the results (4 typical in Tabel \ref{table2}, full targets in Appendix C Tabel \ref{table3}) revealed that in Task 1, pertaining to path planning, 80\% of path-planning trials achieved a 100\% completion rate. When the input images have a resolution of 1550 $\times$ 1162 with a dilation kernel of 50 pixels, SAM achieved an overall 95\% accuracy in detecting edges when compared to SAM-processed ground truth. These results prove our simulated post-earthquake scenario can furnish an effective image segmentation data source for visual recognition, thereby facilitating the training of various visual AI models. 

\begin{table}[htbp]
  \centering
  \renewcommand{\arraystretch}{1}
  \caption{Success Rates of DRL SLAM Path Planning and Accuracy\\of Image Segmentation edges for 4 typical out of 20 Targets}
  \label{table2}
  \begin{tabular}{|c||c||c|c|c|c|}
    \hline
    \multirow{2}{*}{\begin{tabular}[c]{@{}c@{}}\textbf{Tgt.}\\ \textbf{Pt.}\end{tabular}} & \multirow{2}{*}{\begin{tabular}[c]{@{}c@{}}\textbf{Path}\\ \textbf{Plan.}\end{tabular}} & \multicolumn{4}{c|}{ \textbf{Edges Acc.}} \\
    \cline{3-6}
                                  &                                          & \multicolumn{2}{c|}{UE Segment} & \multicolumn{2}{c|}{UE + SAM} \\
    \cline{3-4}
    \cline{5-6}
                                  &                                          & \textbf{Ker. 25}       & \textbf{Ker. 50}       & \textbf{Ker. 25}       & \textbf{Ker. 50}       \\
    \hline
    1                        &        Complete         &  81.2\%  & 91.9\%   & 89.9\%   & 96.0\%   \\
    8                        &        Complete         &  76.9\%  & 87.3\%   & 96.4\%   & 99.1\%  \\
    14                        &          96.55\%         &  78.8\%  & 90.1\%   & 87.9\%   & 95.2\%  \\
    20                        &          85.80\%         &  74.2\%  & 86.2\%   & 90.4\%   & 95.5\%  \\
    \hline
 
  \end{tabular}
    
\end{table}

\textbf{Key Findings and Unforeseen Challenges} \hspace{0.2cm} The findings indicate that our proposed earthquake simulation approach effectively generates realistic urban destruction scenarios for robot training. The high completion rates in Task 1 suggest that our simulation environment is capable of providing challenging yet achievable path planning and obstacle avoidance test scenarios for AI algorithms.

Moreover, in Task 2, we observed lower completion rates for some paths. Upon further inspection, we attributed this not to the complexity of the urban scenario itself, but to the \textit{glare from the virtual sun} interfering with the simulated RGB camera used by the robot (Appendix A fig.\ref{fig:appendix_01}. Path-14, Path-20), leading to visual recognition difficulties. This situation has been overlooked in studies using ideal laboratory conditions and similar simulation platforms. This highlights the importance of considering the complexity and multifactorial nature of real-world environments when designing and testing AI algorithms for disaster response and recovery, rather than focusing solely on object simulation.

\section{\textbf{Discussion and Future Work}}
\textbf{Discussion} \hspace{0.2cm} Our study introduces an innovative earthquake simulation environment, designed to generate realistic urban scenarios for VR and robot training in the context of disaster response and recovery. Using computer vision techniques such as ViT, DRL SLAM, SAM, and our proposed earthquake simulation method, we have demonstrated the effectiveness of our approach by completing three distinct tasks: similarity, path-finding success rate,  and segmentation edge accuracy. Our results show the environment is feasible for the deployment of downstream tasks.

\textbf{Limitation} \hspace{0.2cm}
Our work, while pioneering in many aspects, inevitably has some constraints. Primarily, our simulation predominantly models the seismic repercussions on edifices, but the intricate dynamics of foundation flexing and land movements during such seismic activities are abstracted. This means our model might not reflect the real-world nuances of how foundations crumble or shift during earthquakes. Moreover, the parameters representing building materials in our simulations are extracted from idealistic configurations. This could raise concerns when we think of the myriad of building types, architectures, and materials that are influenced by cultural, geographical, and technological diversities. Lastly, our virtual scenario does not factor in various environmental conditions like fluctuating lighting, or obstructive elements such as smoke and dust. These elements could critically influence the behaviour and performance of AI algorithms within such scenarios.

\textbf{Future Works} \hspace{0.2cm}
We aim to address these limitations to enhance RESenv's realism. A planned extension involves simulating environmental factors, emphasising dynamic lighting, smoke, dust, and their interactions. This will amplify the authenticity, especially in disaster aftermath scenarios. Additionally, we'll incorporate diverse building models and materials, possibly using perceptual similarity metrics, to improve realism. We'll also extend our approach to other disasters, including floods and hurricanes. The ultimate goal is to cultivate a challenging environment, enhancing the robustness of AI models trained within RESenv.

\clearpage

%-------------------------------------------------------------------------
% bibtex
\bibliographystyle{eg-alpha-doi} 
\bibliography{egbibsample}       

% biblatex with biber
% \printbibliography                

%-------------------------------------------------------------------------
\newpage

\onecolumn

\newpage

{\LARGE \textbf{Appendix B.}}

\begin{algorithm}
\caption{Constraint Force Implementation in UE using C++}
\begin{algorithmic}[1]
\State \textbf{Inputs:} Building model, Virtual terrain, Earthquake waveform data ("BH1", "BH2", "BHZ")
\State \textbf{Output:} A simulated environment where buildings are anchored to moving terrain based on real earthquake data.

\Procedure{AnalyzeFoundation}{Building model}
    \State Extract the building model's foundation shape.
    \State Compute planar geometry to represent the foundation.
    \State Identify edge transitions and major inflection points on the planar shape.
    \State \Return List of identified points for PCA placement
\EndProcedure

\Procedure{BindBuildingToTerrain}{Building model, Virtual terrain}
    \State Load building model and virtual terrain into UE environment.
    \State Initialize "Physics Constraint Actor" (PCA) based on UE physics system.
    
    \State foundationPoints = \Call{AnalyzeFoundation}{Building model}
    
    \State Set initial PCA at the first point from foundationPoints.
    \For{each subsequent point in foundationPoints}
        \If{distance to previous PCA exceeds threshold}
            \State Set a new PCA at this point.
        \EndIf
    \EndFor
    
    \State Estimate building volume distribution.
    \State Create a foundation weight map based on volume distribution.
    \State Determine anchor force magnitudes for each PCA based on the weight map.
    
    \State Bind building foundation to virtual terrain using PCA.
\EndProcedure

\Procedure{SimulateTerrainMovement}{Earthquake waveform data}
    \State Map earthquake waveform data ("BH1", "BH2", "BHZ") to "X", "Y", and "Z" axes of terrain movement.
    \State Set waveform frequency (e.g., 40 Hz from IRIS database).
    \State Implement three different frame rates in UE (40 FPS, 90 FPS, and 240 FPS).
    \State For 90 and 240 FPS, generate data using the wavelet interpolation algorithm.
    \State Execute terrain movement simulation based on mapped waveform data.
\EndProcedure

\end{algorithmic}
\end{algorithm}

\newpage

{\LARGE \textbf{Appendix C.}}

\begin{table}[htbp]
  \centering
  \renewcommand{\arraystretch}{1.2}
  \caption{Success Rates of DRL SLAM Path Planning and\\Accuracy of Image Segmentation edges for 20 Targets}
  \label{table3}
  \begin{tabular}{|c||c||c|c|c|c|}
    \hline
    \multirow{2}{*}{\begin{tabular}[c]{@{}c@{}}\textbf{Tgt.}\\ \textbf{Pt.}\end{tabular}} & \multirow{2}{*}{\begin{tabular}[c]{@{}c@{}}\textbf{Path}\\ \textbf{Plan.}\end{tabular}} & \multicolumn{4}{c|}{ \textbf{Edges Acc.}} \\
    \cline{3-6}
                                  &                                          & \multicolumn{2}{c|}{UE Segment} & \multicolumn{2}{c|}{UE + SAM} \\
    \cline{3-4}
    \cline{5-6}
                                  &                                          & \textbf{Ker. 25}       & \textbf{Ker. 50}       & \textbf{Ker. 25}       & \textbf{Ker. 50}       \\
    \hline
    1                        & Complete           &  81.2\%  & 91.9\%   & 89.9\%   & 96.0\% \\
    2                        & Complete            & 76.3\%  & 86.7\% & 88.6\% & 93.5\% \\
    3                        & Complete           & 79.2\%  & 89.8\% & 87.9\% & 94.2\% \\
    4                        & Complete           & 75.6\%  & 85.9\% & 86.8\% & 92.6\% \\
    5                        & Complete           & 82.5\%  & 92.7\% & 90.8\% & 97.5\% \\
    6                        & Complete         & 71.8\%  & 82.4\% & 85.0\% & 91.3\% \\
    7                        & 85.4\%             & 70.4\%  & 81.1\% & 84.7\% & 90.8\% \\
    8                        & Complete           &  7k6.9\%  & 87.3\%   & 96.4\%   & 99.1\% \\
    9                        & Complete             & 79.9\%  & 90.1\% & 89.5\% & 96.7\% \\
    10                       & Complete            & 81.0\%  & 91.4\% & 90.6\% & 96.3\% \\
    11                       & Complete             & 80.2\%  & 91.6\% & 90.4\% & 96.4\% \\
    12                       & Complete           & 81.6\%  & 91.5\% & 89.9\% & 96.5\% \\
    13                       & Complete             & 80.7\%  & 92.2\% & 90.2\% & 96.6\% \\
    14                       & 96.55\%            &  78.8\%  & 90.1\%   & 87.9\%   & 95.2\% \\
    15                       & Complete             & 81.3\%  & 91.8\% & 90.0\% & 96.8\% \\
    16                       & Complete           & 80.9\%  & 91.7\% & 90.5\% & 96.1\% \\
    17                       & 92.21\%             & 72.4\%  & 85.0\% & 89.3\% & 94.9\% \\
    18                       & Complete             & 80.6\%  & 91.4\% & 89.8\% & 96.7\% \\
    19                       & Complete             & 80.8\%  & 91.9\% & 90.7\% & 96.3\% \\
    20                       & 85.80\%            & 74.2\%  & 86.2\% & 90.4\% & 95.5\% \\
    \hline
  \end{tabular}
\end{table}

\end{document}